\documentclass[a4paper,12pt]{article}
\usepackage{amsmath,amsfonts,amssymb,epsfig}
\usepackage{fullpage}
\usepackage{natbib}

\usepackage{color}

\newcommand{\be}{\begin{equation}}
\newcommand{\en}{\end{equation}}

\renewcommand{\vec}[1]{\boldsymbol{#1}}
\def \bm#1{\mbox{\boldmath{$#1$}}}   

\def \tr{\mbox{tr\hskip 1pt}}

\begin{document}


\title{Large acoustoelastic effect  }


\author{Zaki Abiza$^a$, Michel Destrade$^{b,c}$, Ray W. Ogden$^d$ \\[12pt]
$^a$Ecole Nationale Sup\'erieure  d'Electronique, Informatique,\\ T\'el\'ecommunications, 
Math\'ematiques et M\'ecanique de Bordeaux,\\
1 avenue du Docteur Albert Schweitzer \\B.P. 99,
33402 Talence Cedex,
France\\[12pt]
$^b$School of Mathematics, Statistics and Applied Mathematics,  \\
   National University of Ireland Galway, \\University Road, Galway, Ireland \\[12pt]
$^c$School of Mechanical and Materials Engineering, \\
University College Dublin, Belfield, Dublin 4, Ireland\\[12pt]
$^d$School of Engineering,\\
    University of Aberdeen,  \\
    King's College, Aberdeen AB24 3UE, United Kingdom}

\date{}

\maketitle


\begin{abstract}

Classical acoustoelasticity couples small-amplitude elastic wave propagation to an infinitesimal pre-deformation, in order to reveal and evaluate non-destructively third-order elasticity constants.
Here, we see that acoustoelasticity can be also be used to determine fourth-order constants, simply by coupling a small-amplitude wave with a small-but-finite pre-deformation. 
We present results for compressible weakly nonlinear elasticity, we make a link with the historical results of Bridgman on the physics of high pressures, and we show how  to determine ``$D$", the so-called fourth-order elasticity constant of soft (incompressible, isotropic) solids by using infinitesimal waves.

\end{abstract}


\newpage

\numberwithin{equation}{section}


\section{Introduction\label{Introduction}}


Acoustoelasticity is now a well established experimental technique used for the non-destructive measurement of third-order elasticity (TOE) constants of solids, and its principles can be found in standard handbooks of physical acoustics, such as \cite{PaSa84, KiSa01}.
The underlying theory, however, is quite intricate, and over the years several authors have produced different and irreconcilable expressions for the shift experienced by the wave speed when elastic wave propagation is coupled to a pre-strain.
A common mistake found in the literature consists of the implicit assumption that since a small pre-strain and a small-amplitude wave are described by linearized equations, they can be superposed linearly. With that point of view, the coupling between the two phenomena is of higher order and can be neglected in the first approximation. The flaw in that reasoning is simply that these phenomena are \emph{successive}, not linearly superposed. Experiencing an infinitesimal pre-strain from a stress-free configuration is a linear process; propagating an infinitesimal wave in a stress-free configuration is another linear process.
But in acoustoelasticity, the wave travels in a pre-stressed, not a stress-free solid, and the laws of linear elastodynamics must be adapted to reflect this fact.
The final outcome of this analysis is that not only second-order, but also TOE constants appear in the expression for the speed of an acoustoelastic wave.

For example, a longitudinal wave travelling in a solid subject to a hydrostatic stress $\sigma \vec I$ will propagate with the speed $v_{0L}$ given by
\begin{equation} \label{rhov2}
\rho v_{0L}^2  = \lambda + 2 \mu -\frac{7 \lambda + 10 \mu + 2A + 10B + 6C}{3\lambda + 2\mu} \sigma,
\end{equation}
where $\rho$ is the mass density \emph{in the unstressed configuration}, $\lambda$ and $\mu$ are the (second-order) Lam\'e coefficients, and $A$, $B$, $C$ are the (third-order) Landau coefficients \citep{LaLi86}. This  equation was correctly established as early as 1925 by Brillouin and confirmed in 1953 by Hughes and Kelly, although several erroneous expressions have appeared in between and since (see, e.g., \cite{Birc38, Tang67}).

From our contemporary perspective, the easiest way to re-establish and further this expression is to rely on the modern theory of incremental (also known as small-on-large) elasticity, which presents compact expressions for the tensor of \emph{instantaneous elastic moduli}, $\vec{\mathcal{A}}_0$, given by its components
\begin{equation}
\left\{
\begin{aligned}
&\mathcal{A}_{0iijj} = J^{-1}\lambda_i \lambda_j W_{ij} , &\\[1ex]
& \mathcal{A}_{0ijij} = J^{-1}\frac{\lambda_i W_i - \lambda_j W_j}{\lambda_i^2 -\lambda_j ^2} \lambda_i^2, & i \ne j, \; \lambda_i \ne \lambda_j ,\\[1ex]
& \mathcal{A}_{0ijji} = J^{-1}\frac{\lambda_j W_i - \lambda_i W_j}{\lambda_i^2 -\lambda_j ^2} \lambda_i \lambda_j, & i \ne j, \; \lambda_i \ne \lambda_j ,\\[1ex]
& \mathcal{A}_{0ijij} = J^{-1} (\lambda_i^2 W_{ii} - \lambda_i\lambda_jW_{ij} + \lambda_i W_i)/2, &  i \ne j, \; \lambda_i = \lambda_j,\\[1ex]
& \mathcal{A}_{0ijji} = J^{-1} (\lambda_i^2 W_{ii} - \lambda_i\lambda_jW_{ij} - \lambda_i W_i)/2, &  i \ne j, \; \lambda_i = \lambda_j,  \end{aligned}
\right. \label{Aoijkl}
\end{equation}
in the coordinate system aligned with the principal axes of pre-strain, which, since the material is isotropic, coincide with the principal axes of pre-stress.
Here, $W$  is the strain-energy density per unit volume, which is a symmetric function of the principal stretches $\lambda_1,\lambda_2,\lambda_3$ of the deformation, $J = \lambda_1\lambda_2\lambda_3$, $W_i=\partial W/\partial\lambda_i$ and $W_{ij}=\partial^2W/\partial\lambda_i\partial\lambda_j,\,i,j\in\{1,2,3\}$; see, for example,  \cite{Ogde97,Ogde07} for details.  The corresponding principal Cauchy stresses are given by
$\sigma_i=J^{-1}\lambda_iW_i,\,i=1,2,3$.

It is worth noting here for later reference that it follows from \eqref{Aoijkl}$_2$ that
\begin{equation}
\mathcal{A}_{0ijij} -\mathcal{A}_{0jiji}=\sigma_i-\sigma_j,\quad i\neq j.\label{calAdifference} 
\end{equation}

A small-amplitude body wave may travel at speed $v$ in the direction of the unit vector $\vec{n}$ with polarization in the direction of a unit vector $\vec{m}$ provided the eigenvalue problem
\begin{equation} \label{rhov2m}
\vec{Q}(\vec{n}) \vec{m} = \rho v^2 \vec{m}
\end{equation}
is solved with $\rho v^2 >0$,
where $\vec{Q}(\vec{n})$ is the \emph{acoustic tensor}, which has components
\begin{equation} \label{Qij}
Q_{ij}(\vec{n}) = J\mathcal{A}_{0piqj} n_p n_q.
\end{equation}
Note that sometimes the acoustic tensor is defined by \eqref{Qij} \emph{without} the factor $J$, in which case the $\rho$ in \eqref{rhov2m} would be replaced by the density in the deformed configuration, i.e. $J^{-1}\rho$.

Notice that in this theory the amplitude of the acoustic wave is infinitesimal, but that there is \emph{no restriction} on the choice of $W$ or on the magnitude of the underlying pre-strain because we are in the context of \emph{finite} nonlinear elasticity.
It follows that the expressions  can be specialized in various ways, in particular to \emph{weakly} nonlinear elasticity (where $W$ is expanded in terms of some measure of strain) and to special pre-strains.

Hence, to access the TOE constants, we take $W$ as
\begin{equation} \label{energy1}
W = W_\text{TOE} \equiv \frac{\lambda}{2}I_{1}^2 + \mu I_2 + \frac{A}{3} I_3 + B I_1 I_2 + \frac{C}{3} I_{1}^3,
\end{equation}
where $I_n = \tr(\vec{E}^n),\,n = 1, 2, 3$, are invariants of the Green--Lagrange strain tensor $\vec{E}$.
The components \eqref{Aoijkl} can then be expanded to the first order in terms of, for example, the volume change $\varepsilon\equiv J - 1$ in the case of a hydrostatic pressure, or of the elongation $e_1\equiv\lambda_1- 1$ in the case of uniaxial tension.

To access fourth-order elasticity (FOE) constants, we take $W$ as
\begin{equation} \label{energy2}
W = W_\text{FOE} \equiv W_\text{TOE} + E I_1 I_3  + F I_1^2 I_2 + G I_2^2 + H I_1^4,
\end{equation}
where $E$, $F$, $G$, $H$ are the FOE constants and we push the expansions of the wave speed up to the next order in the strain.
This is what we refer to as the \emph{large acoustoelastic effect}. In fact, the main purpose of this investigation is to provide
a theoretical backdrop to the current drive to determine experimentally the FOE constants of soft solids such as isotropic tissues
and gels in order to improve acoustic imaging resolution (see, e.g., \cite{HaIZ04, HaIZ04B, HaIZ07, Genn07, Reni07, JCGB07, Reni08b, Reni08a, Miro09}).

Because soft solids are often treated as \emph{incompressible}, so that the constraint $J \equiv 1$ must be satisfied, we shall also consider the expressions
for the TOE and FOE strain-energy densities in their incompressible specializations, specifically
\begin{equation} \label{energytoe}
W = \mu I_2 + \dfrac{A}{3}I_3
\end{equation}
for TOE incompressibility and
\begin{equation} \label{energyfoe}
W = \mu I_2 + \frac{A}{3}I_3 + D I_2^2
\end{equation}
for FOE incompressibility.
Hence, in the transition from compressible to incompressible elasticity, the number of second-order constants goes from two to one, of third-order constants from three to one, and of fourth-order constants from four to one.  This was established by Hamilton et al.  in 2004, although it can be traced back to Ogden  in 1974 and even earlier, to Bland  in 1969. 
Specifically, in that transition we note that the elasticity constants behave as \citep{DeOg10}
\begin{equation} \label{incompr2}
\lambda \rightarrow \infty, \quad \mu \rightarrow E_Y/3,
\end{equation}
for the second-order constants, where $E_Y$ is the (finite) Young's modulus,
\begin{equation}
A/\mu = \mathcal{O}(1), \quad B/\mu=\mathcal{O}(\lambda/\mu), \quad C/\mu = \mathcal{O}(\lambda^2/\mu^2)
\end{equation}
for the third-order constants,
\begin{equation} \label{incompr3}
E/\mu = \mathcal{O}(\lambda/\mu),  \quad
F/\mu = \mathcal{O}(\lambda^2/\mu^2), \quad
G/\mu = \mathcal{O}(\lambda/\mu), \quad
H/\mu = \mathcal{O}(\lambda^3/\mu^3),
\end{equation}
for the fourth-order constants, and $(C+F)/\mu=\mathcal{O}(\lambda/\mu)$ for a combination of second- and third-order constants.
Some constants have the explicit limiting behaviour
\begin{equation} \label{incompr4}
B/\lambda \rightarrow -1, \quad
E/\lambda \rightarrow 4/3, \quad
G/\lambda \rightarrow 1/2, \quad
B + G + \lambda/2 \rightarrow D,
\end{equation}
where $D$ remains a finite quantity, of the same order of magnitude as $\mu$; see \cite{DeOg10} for analysis of the behaviour of $A$, $B$, \dots, $H$ in the incompressible limit and the connection with $D$. Note that the limiting behaviour  for $E$ in terms of the initial Poisson's ratio $\nu$ and Young's modulus $E_Y$ is thus $(1-2\nu) E \rightarrow 4E_Y/9$. This was shown in \cite{DeOg10}, but mistyped in equations (49) and (89) therein as $(1-2\nu) E \rightarrow 4E_Y/3$.

In this paper, we treat in turn the case of hydrostatic pre-stress (Section \ref{Hydrostatic pressure}) and of uniaxial pre-stress (Section
\ref{Uniaxial tension}), and we provide expansions of the body wave speeds up to the second order in the pre-strain, and also in the pre-stress, for compressible solids and in the relevant incompressible limits.


\section{Hydrostatic pressure\label{Hydrostatic pressure}}


Consider first a cuboidal sample of a compressible solid with sides of lengths $(L_1,L_2,L_3)$ in its (unstressed) reference configuration.  We define the reference geometry in terms of Cartesian coordinates $(X_1,X_2,X_3)$ by $0 \leq X_i \leq  L_i, \, i = 1, 2, 3 $.  The material is then subject to a pure homogeneous strain $x_1=\lambda_1X_1,x_2=\lambda_2X_2,x_3=\lambda_3X_3$ and deformed into the cuboid $ 0 \leq x_i \leq  l_i, \,i = 1, 2, 3$, where $(x_1,x_2,x_3)$ are the Cartesian coordinates in the deformed configuration, and the constants $\lambda_1,\lambda_2,\lambda_3$ are the principal stretches of the deformation.
We now specialize the deformation to a pure dilatation so that $\lambda_1 = \lambda_2 = \lambda_3=J^{1/3}$.  Since the material is isotropic the Cauchy stress $\vec \sigma$ is spherical, $\vec \sigma = \sigma \vec I$ say, with $\sigma >0\,(<0)$ corresponding to hydrostatic tension (pressure).

The pre-stress is computed as $\sigma = J^{-1}\lambda_1W_1$ or, more conveniently, as $\sigma = \hat W'(J)$, where $\hat W(J) \equiv W(J^{1/3}, J^{1/3}, J^{1/3})$.
Up to the second order in the volume change $\varepsilon \equiv J - 1$, we find
\begin{equation} \label{sigmaeps}
\sigma =\kappa  \varepsilon + \tfrac{1}{2} \hat W'''(1) \varepsilon^2,
\end{equation}
where $\kappa = \hat W''(1) = \lambda + 2\mu/3$ is the infinitesimal bulk modulus, and
\begin{equation} \label{W'''}
\hat W'''(1) = -\kappa + \dfrac{2}{9}(A + 9B + 9C).
\end{equation}
(It is a simple matter to check that the expansion \eqref{sigmaeps} is equivalent to one first established by  \cite{Murn51}.)
Conversely, the volume change is expressed in terms of the hydrostatic stress as
\begin{equation} \label{epsilonsigm}
\varepsilon =  \dfrac{1}{\kappa}  \sigma -  \dfrac{\hat W'''(1)}{2\kappa^3}\sigma^2.
\end{equation}

Here the pre-stress does not generate preferred directions and the solid remains isotropic.
It follows that two waves may propagate in any direction, one longitudinal, with speed $v_{L}$, and one transverse, with speed $v_{T}$, given by
\begin{equation}
\rho v_{L}^2 = J\mathcal{A}_{01111}, \quad
\rho v_{T}^2 = J\mathcal{A}_{01212},
\end{equation}
where these quantities are computed from the formulas \eqref{Aoijkl}.
Expanding in terms of the volume change, we obtain
\begin{equation} \label{hydro-e}
\rho v_{L}^2 =  \lambda + 2 \mu + a_{L} \varepsilon + b_{L} \varepsilon^2,
 \quad
 \rho v_{T}^2 =  \mu + a_{T} \varepsilon + b_{T}  \varepsilon^2,
\end{equation}
where
\begin{equation}
a_{L} = \frac{7}{3} \lambda + \frac{10}{3} \mu+  \frac{2}{3}A  + \frac{10}{3}B + 2 C,
\quad
a_{T} =  \lambda + 2 \mu + \frac{1}{3}A +   B,
\end{equation}
are the coefficients for the classical (linear) acoustoelastic effect, and
\begin{eqnarray}
 b_{L}& =&\frac{13}{18} \lambda + \frac{7}{9} \mu+  \frac{8}{9}A  + \frac{44}{9}B + \frac{10}{3} C + \frac{8}{3} E + \frac{16}{3} F + \frac{20}{9} G + 12H, \notag \\[1ex]
b_{T} &=&\frac{1}{2} \lambda + \frac{5}{9} \mu+  \frac{1}{2} A  + \frac{13}{6} B + C + E +  F + \frac{2}{3} G
\end{eqnarray}
are the coefficients of the large (quadratic) acoustoelastic effect.

Alternatively, we may use \eqref{epsilonsigm} to express the wave speeds in terms of the pre-stress rather than the pre-strain, as
\begin{equation} \label{hydro-s}
\rho v_{L}^2 =  \lambda + 2 \mu + c_{L}\sigma   + d_{L} \sigma^2,\quad \rho v_{T}^2 =  \mu + c_{T} \sigma + d_{T} \sigma^2,
\end{equation}
thus recovering the classical formulas 
\begin{equation}
c_{L} = \frac{a_{L}}{\kappa},\quad c_{T} =  \frac{a_{T}}{\kappa},
\end{equation}
for the (linear) acoustoelastic effect \citep{HuKe53}, and establishing the formulas
\begin{equation}
d_{L} = \frac{b_{L}}{\kappa^2} - \dfrac{a_{L} \hat W'''(1)}{2 \kappa^3},\quad d_{T} =  \frac{b_{T}}{\kappa^2} - \dfrac{a_{T} \hat W'''(1)}{2 \kappa^3}
\end{equation}
for the large (quadratic) acoustoelastic effect, where $\hat W'''(1)$ is given in \eqref{W'''}.

\begin{figure}
  \begin{center}
\epsfig{figure=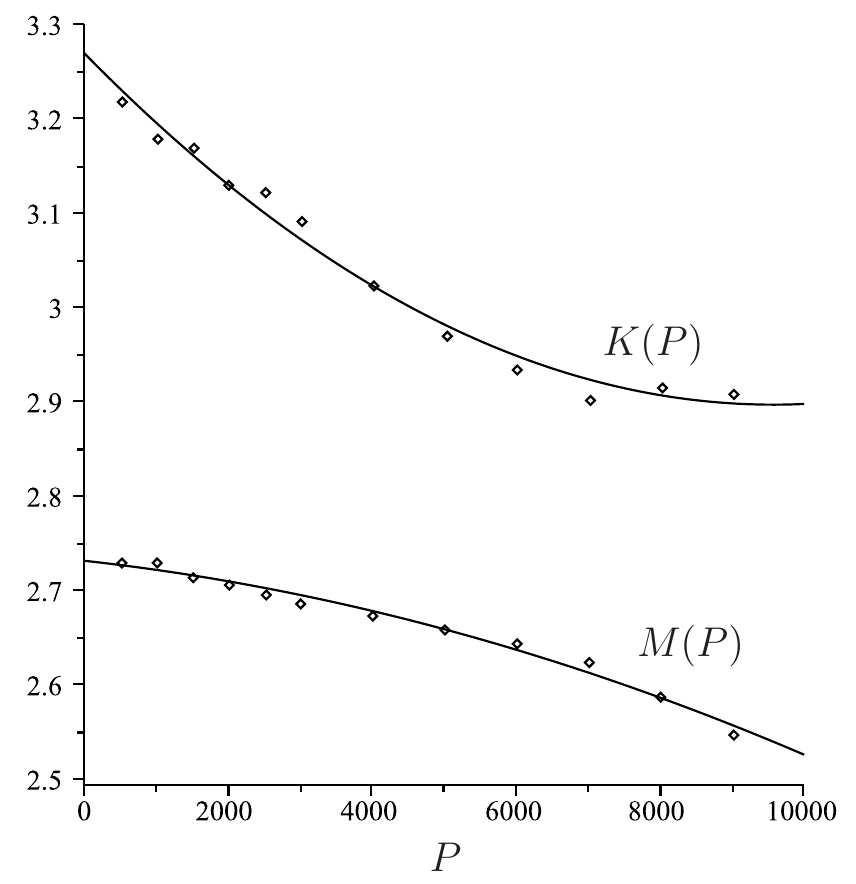, width=.5\textwidth}
\end{center}
\caption{Fit of the data measured by  \cite{HuKe53} for pyrex subjected to hydrostatic pressure $P$ (in bars) based on the equations \eqref{hydro-s} and \eqref{MK} with $\sigma=-P$ describing the large acoustoelastic effect: $M$ and $K$ are expressed in units of $10^5$ bars. } \label{fig:hydro}
\end{figure}

For an application of these results, we turn to the classical data of \cite{HuKe53}.
They performed acoustoelastic experiments and plotted the stress-dependent shear and bulk moduli, defined as
\begin{equation} \label{MK}
M \equiv \rho v^2_{T}, \quad K \equiv \rho v^2_{L} - \frac{4}{3} \rho v^2_{T},
\end{equation}
against the hydrostatic pressure $P=-\sigma$ for polysterene and for pyrex.
For the former, the variations are clearly linear, and nothing would be gained by including FOE effects.
For the latter, there is a marked departure from the linear acoustoelastic effect at high pressures, and we thus use \eqref{hydro-s} to obtain a better fit to the data. By digitizing the data and performing a standard least square optimization, we found that for pyrex,
\begin{eqnarray}
M(P)& = & 2.73 - 8.45 \times 10^{-6} P - 1.21 \times 10^{-9}P^2, \notag \\[1ex]
K(P) & = & 3.27 - 7.80 \times 10^{-5}P + 4.07 \times 10^{-9}P^2,
\end{eqnarray}
where $M$ and $K$ are expressed in units of $10^5$ bars and $P$ in bars. The fit to the data is shown in Fig. \ref{fig:hydro}.

According to \eqref{hydro-s}, the quantities \eqref{MK} give access to $\lambda$ and $\mu$, to two linear combinations of the third-order constants, and to two linear combinations of the fourth-order constants. 
Specifically, we may solve \eqref{MK} to find the Lam\'e constants as
\begin{equation}
\lambda = K(0) - \frac{2}{3}M(0), \quad \mu = M(0).
\end{equation}
Similarly, we find two linearly independent combinations of the three third-order constants: 
\begin{equation}
 A+3B = -3 K(0) \left[ 1 + M'(0) \right] - 4 M(0), \quad A + 9B + 9C = - 9K(0)  \left[1+K'(0) \right]/2.
\end{equation}
Finally, we also find two combinations of the fourth-order constants:
\begin{align}
& E + F  + \dfrac{2}{3}G = \dfrac{7}{6}K(0) + \dfrac{4}{3}M(0)
\notag \\[4pt]
& \phantom{ E + F  + \dfrac{2}{3}G = \dfrac{7}{6}K(0)}
 + \dfrac{1}{2}K(0)\left[ K'(0) + \dfrac{13}{3}M'(0) + K'(0)M'(0) + K(0)M''(0) \right],\notag \\[4pt]
& E + 3F + G + 9H =  \dfrac{17}{24}K(0) + \dfrac{3}{2}K(0)\left[K'(0) + \dfrac{1}{4}\left(K'(0) \right)^2 + \dfrac{1}{4}K(0)K''(0)\right].
\end{align}
Hence, for the pyrex data:
\begin{align}
& \lambda = 1.45 \times 10^5 \text{ bars}, &&
\mu = 2.73 \times 10^5 \text{ bars}, \notag \\
& A + 3B = -1.24 \times 10^6 \text{ bars}, &&
A+9B + 9C = 1.00 \times 10^7 \text{ bars}, \notag \\
& E + F  + 2G/3 = -1.30 \times 10^{7} \text{ bars}, &&
E + 3F + G + 9H  = 3.65 \times 10^{7} \text{ bars}.
\end{align}
Clearly,  the data of \cite{HuKe53} are not sufficient to determine all the third- and fourth-order constants.  An additional relation is needed to determine experimentally all the third-order constants, and a further two relations to determine the fourth-order constants.  These may be provided from, for example, wave speed measurements in uniaxially deformed samples, as we show in the following section.

We conclude this section by evoking the work of P.W. Bridgman on ``The Physics of High Pressures'' \citep{Brid45}, which won him the 1946 Nobel Prize. 
He carried out countless high pressure measurements on solids, and fitted his data with a quadratic equation for the volume change, in the form $\Delta V/V_0 = - a P + b P^2$. 
Direct comparison of this experimental law with \eqref{epsilonsigm} reveals the identifications $a = 1/\kappa$ and $b = - \hat W'''(1)/(2\kappa^3)$.
Further, acoustoelastic measurements give direct access to these constants, because
$\kappa = K(0)$ and $\hat W'''(1) = -K(0)[2+K'(0)]$.
In Bridgman's experiments, very high pressures are required to obtain second-order deformations; however, his coefficients $a$ and $b$ can also be determined by pressurizing the sample only  linearly, and then measuring the speeds of transverse and longitudinal waves. 
In fact, had Bridgman been able to propagate such waves in his finitely deformed samples he would have had access to the third term in the expansion,  $\Delta V/V_0 = - a P + b P^2 - cP^3$ say, because the continuation of \eqref{epsilonsigm} is
\begin{equation}
\varepsilon = \dfrac{\Delta V}{V_0} =  -\dfrac{1}{\kappa} P - \dfrac{1}{2\kappa^3}\hat W'''(1)P^2   - \dfrac{1}{6\kappa^5}\left[ 3 \left(\hat W'''(1)\right)^2 - \kappa \hat W^{(4)}(1)\right]P^3,
\end{equation}
where $\hat W^{(4)}(1)$ is found to be expressible as
\begin{equation}
\hat W^{(4)}(1) = \dfrac{19}{9}\kappa - \dfrac{4}{9}(A+9B+9C) + \dfrac{8}{3}(E+3F+G+9H),
\end{equation}
in which each term can be determined experimentally using the equations of large acoustoelasticity above.


\section{Uniaxial tension\label{Uniaxial tension}}


We again consider the cuboidal sample as in Section \ref{Hydrostatic pressure}, but now the cuboid is subject to a uniaxial tension in the $X_1$ direction so that it deforms homogeneously with elongation $e_1 ={l_1}/{L_1} - 1 = \lambda_1 -1$ in the $x_1$-direction.
By symmetry it contracts laterally and equibiaxially with elongation $e_3=e_2 = {l_2}/{L_2} - 1 = \lambda_2 -1$. The nominal stress (axial force per unit reference area) is $J\sigma_1/\lambda_1=\sigma_1 \lambda_2^2 $.  

The condition that the lateral faces are free of traction is
\begin{equation} \label{sigma}
\sigma_3\equiv \sigma_2 = J^{-1}\lambda_2 W_2 = 0,
\end{equation}
which determines the extent of the lateral contraction $ -1< e_2 \leq 0$ (assuming that $\lambda_2<1$) and can be written in terms of $e_1$, up to second order, as
\begin{equation}\label{elong2}
e_2 =  -\alpha e_1-\beta e_1^2,
\end{equation}
where
\begin{equation}
\alpha=\dfrac{\lambda}{2(\lambda+\mu)},\quad \beta=\frac{3\kappa \lambda}{8(\lambda+\mu)^2}+ \frac{\lambda^2 A}{8(\lambda+\mu)^3}
 + \left[\frac{\lambda(\lambda-2\mu)}{2(\lambda+\mu)^2}+1 \right]  \frac{B}{2(\lambda+\mu)}  + \frac{ \mu^2 C}{2(\lambda+\mu)^3}, \label{alphabeta}
\end{equation}
and again $\kappa = \lambda + 2\mu/3$ is the infinitesimal bulk modulus.
Note that \eqref{elong2} collapses to $ e_2 = - e_1/2 +  3e_1^2/8 $ in the incompressible limits \eqref{incompr2} and \eqref{incompr3}, as expected from the expansion to second order of the connection $\lambda_2 = \lambda_1^{-1/2}$ between the stretch ratios of an incompressible solid in uniaxial tension.

Substituting \eqref{elong2} with \eqref{alphabeta} into the expression $\sigma_1 = J^{-1} \lambda_1 W_1$ for the uniaxial stress, we obtain the relation between the pre-stress and the pre-strain, up to the second order, as
\begin{equation} \label{sigma1}
\sigma_1 = \dfrac{3\kappa \mu}{\lambda+\mu} e_1+  \gamma e_1^2,
\end{equation}
where
\begin{eqnarray}
\gamma = \frac{3\kappa\mu\left(5\lambda + 3 \mu\right)}{2\left(\lambda + \mu\right)^2}+ \left[1 - \frac{\lambda^3}{4\left(\lambda + \mu\right)^3}\right] A
+ \frac{3\mu\left(3\lambda^2 + 4 \lambda\mu + 2 \mu^2\right)}{2\left(\lambda + \mu\right)^3} B + \frac{\mu^3}{\left(\lambda + \mu\right)^3} C.\quad
\end{eqnarray}
(Note that although $3\kappa\mu/(\lambda+\mu)=E_Y$, the infinitesimal Young's modulus, we shall not use $E_Y$ hereon.)
Conversely,
\begin{equation} \label{elong1}
 e_1 = \dfrac{\lambda+\mu}{3 \kappa\mu} \sigma_1 - \left(\dfrac{\lambda+\mu}{3 \kappa\mu}\right)^3\gamma \sigma_1^2.
\end{equation}
In the incompressible limits \eqref{incompr2} and \eqref{incompr3}, the stress--strain and strain--stress relations reduce to
\begin{equation} \label{sig}
\sigma_1 = 3 \mu e_1 + 3\left(\mu + \frac{A}{4}\right)e_1^2, \quad e_1 = \frac{1}{3\mu}\sigma_1 - \frac{1}{9\mu^3}\left(\mu + \frac{A}{4}\right) \sigma_1^2.
\end{equation}

Now we examine the possibility of a small-amplitude body wave travelling in the direction of the unit vector $\vec{n}$ with polarization in the direction of a unit vector $\vec{m}$. For a general direction of propagation, the speeds of the different waves are found as the eigenvalues of the acoustical tensor $\vec Q(\vec n)$ in \eqref{Qij}, i.e. as the roots of a cubic. In this paper, however, we focus on the wave speeds that are relatively simple to determine experimentally. The speeds of non-principal body waves are not easily accessible because they would require transducers to be placed at an angle to the faces of the cuboid, and not in full flat contact, which would lead to additional transmission problems.
This leaves the principal body waves, as explained by  \cite{HuKe53}: first, waves in the direction $x_1$ of the tension, and, second,
principal waves in the $x_2$ and $x_3$ directions, which are equivalent by symmetry.

In the direction of tension there exists a longitudinal wave with speed $v_{11}$, say, and two transverse waves, propagating with the same speed $v_{12}$, say (in fact, these latter two waves may be combined to form a transverse circularly-polarized wave).
Hence, $\vec{n} =  \vec{m} = \vec{e}_1$ for the longitudinal wave, and $\vec{n} = \vec{e}_1$, $\vec{m} = \vec{e}_2$ for the transverse wave, where $\vec{e}_1$ and $\vec{e}_2$ are unit basis vectors corresponding to the coordinates $x_1$ and $x_2$, respectively.
Using \eqref{rhov2m} and \eqref{Qij} we then have
\begin{equation}
\rho v_{11}^2 =J \mathcal{A}_{01111}, \quad  \rho v_{12}^2 =J \mathcal{A}_{01212}.
\end{equation}
Expanding the elastic moduli in terms of the elongation $e_1$, we obtain
\begin{equation} \label{uni-e-1}
\rho v_{11}^2 =  \lambda + 2 \mu + a_{11} e_1 + b_{11} e_1^2,\quad \rho v_{12}^2 =  \mu + a_{12} e_1 + b_{12} e_1^2,
\end{equation}
where
\begin{eqnarray}
a_{11}& =& \dfrac{\mu}{\lambda+\mu}\left(\lambda + 2B + 2C \right) + 2\left(2\lambda+5\mu+A+2B\right),\notag \\[1ex]
a_{12}& = &\dfrac{\mu}{\lambda+\mu}\left[4(\lambda+\mu) + \dfrac{\lambda+2\mu}{4\mu} A +B \right],
\end{eqnarray}
for the classical acoustoelastic effect,
and
\begin{eqnarray} \label{b11}
b_{11}&=&\frac{12 \lambda^2 +51\lambda \mu+34\mu^2}{2(\lambda + \mu)}+ \left[ 10 - \frac{\lambda^3}{4\left(\lambda + \mu \right)^3}\right] A+\frac{3 \left( \lambda + 2 \mu\right) \left(12 \lambda^2+ 21 \lambda \mu + 10 \mu^2  \right)}{2\left(\lambda + \mu \right)^3} B\notag\\[1ex]
&+& \mu \left[ \frac{9}{\lambda + \mu}+  \frac{\mu^2}{\left( \lambda +\mu \right)^3} \right] C
- \frac{\lambda^2 A}{2\left(\lambda + \mu \right)^3} \left( B +C \right)-\frac{3\lambda^2+2\lambda\mu+2\mu^2}{(\lambda+\mu)^3}BC\notag\\[1ex]
&-&\frac{3 \lambda^2 +2 \lambda \mu + 2 \mu^2}{\left(\lambda + \mu \right)^3}B^2-\frac{2\mu^2}{\left(\lambda + \mu \right)^3}C^2+\frac{6\left( \lambda+2\mu \right)}{\lambda+ \mu}E + 3\left( \frac{\lambda+2\mu }{\lambda+ \mu}\right)^2 F\notag\\[1ex]
&+& 2\left[ 6 + \frac{\lambda^2}{\left( \lambda + \mu\right)^2}\right]G + \frac{12 \mu^2}{\left( \lambda + \mu\right)^2}H,
\end{eqnarray}
\begin{eqnarray}\label{b12}
b_{12}& = &6\mu + \frac{9\left(\lambda+2\mu \right)}{8(\lambda+\mu)}A + \frac{9\mu}{2(\lambda+\mu)}B-\frac{\lambda^2}{16(\lambda+\mu)^3}A^2-\frac{3\lambda^2+2\lambda\mu+2\mu^2}{2(\lambda+\mu)^3}B^2\notag\\[1ex]
&-&\frac{5\lambda^2+2\lambda\mu+2\mu^2}{8(\lambda+\mu)^3}AB-\frac{\mu^2}{4(\lambda+\mu)^3}(A+4B)C+\frac{3\mu \left(\lambda+2\mu \right)}{4\left(\lambda+\mu \right) ^2}E+\frac{\mu^2}{\left(\lambda+\mu \right)^2}F\notag\\[1ex]
&+&  \left[ 2 +\frac{\lambda^2}{\left(\lambda+\mu \right)^2}\right]G,
\end{eqnarray}
for the large acoustoelastic effect.

For propagation perpendicular to the direction of tension, we take $\vec{n} = \vec{e}_2$. Then there exists a longitudinal wave, with $\vec{m} = \vec{e}_2$ and speed $v_{22}$, a transverse wave polarized in the direction of tension $\vec{m} = \vec{e}_1$ with speed $v_{21}$,
and a transverse wave polarized perpendicular to the direction of tension, with $\vec{m} = \vec{e}_3$ and speed $v_{23}$.
These speeds are given by
\begin{equation} \label{rhov2LT}
\rho v_{22}^2 = J\mathcal{A}_{02222}, \quad  \rho v_{21}^2 =J \mathcal{A}_{02121}, \quad  \rho v_{23}^2 = J\mathcal{A}_{02323}.
\end{equation}
Expanding the elastic moduli in terms of the elongation $e_1$, we obtain
\begin{equation} \label{uni-e-2}
\rho v_{22}^2 =  \lambda + 2 \mu + a_{22} e_1 + b_{22} e_1^2,\quad \rho v_{2k}^2 =  \mu + a_{2k} e_1 + b_{2k} e_1^2,\quad k=1,3,
\end{equation}
where
\begin{eqnarray}
 a_{22}& =& -\dfrac{2\lambda(\lambda+2\mu)+\lambda A+ 2(\lambda-\mu)B-2\mu C}{\lambda+\mu},\\[1ex]
 a_{21} &=&  \dfrac{(\lambda + 2\mu)(4\mu+A) + 4\mu B}{4(\lambda+\mu)},\\[1ex]
 a_{23}& =& - \dfrac{\lambda(4\mu+A)-2\mu B}{2(\lambda+\mu)},
\end{eqnarray}
are the classical acoustoelastic coefficients, and
\begin{eqnarray} \label{b22}
b_{22} &=&  -\frac{\lambda \mu (\lambda+2\mu )}{( \lambda +\mu )^2} - \frac{\lambda(2\lambda^2-\mu^2)}{2(\lambda+\mu)^3}A-\frac{3\mu(3\lambda^2+3\lambda\mu+\mu^2)}{(\lambda+\mu)^3}B
-\frac{\mu(3\lambda^2+4\lambda\mu+3\mu^2)}{(\lambda+\mu)^3}C\notag\\[1ex]
&-&\frac{\lambda^2}{4(\lambda+\mu)^3}A^2-\frac{2(3\lambda^2+2\lambda\mu+2\mu^2)}{(\lambda+\mu)^3}B^2-\frac{2\mu^2}{(\lambda+\mu)^3}C^2
-\frac{5\lambda^2+2\lambda\mu+2\mu^2}{2(\lambda+\mu)^3}AB\notag\\[1ex]
&-&\frac{3\lambda^2+2\lambda\mu+6\mu^2}{(\lambda+\mu)^3}BC
-\frac{\lambda^2+2\mu^2}{2(\lambda+\mu)^3}AC+\frac{3\lambda(\lambda-2\mu)}{2(\lambda+\mu)^2}E+\frac{3\lambda^2+4\mu^2}{(\lambda+\mu)^2}F\notag\\[1ex]
&+&4\left[1+\frac{\lambda^2}{(\lambda+\mu)^2}\right]G+\frac{12\mu^2}{(\lambda+\mu)^2}H,
\notag \\[4pt]
 b_{21} &= & b_{12} - \gamma -\frac{3\kappa \mu^2}{\left( \lambda +\mu \right)^2},
\notag \\[4pt]
 b_{23} &=& -\frac{\lambda \mu^2}{( \lambda +\mu)^2} + \frac{\lambda(3\lambda^2-\mu^2)}{4(\lambda+\mu)^3}A
 -\frac{3(3\lambda^2+2\lambda\mu+\mu^2)\mu}{2(\lambda+\mu)^3}B-\frac{2\mu^3}{(\lambda+\mu)^3}C-\frac{\lambda^2}{8(\lambda+\mu)^3}A^2\notag\\[1ex]
 &-&\frac{3\lambda^2+2\lambda\mu+2\mu^2}{2(\lambda+\mu)^3}B^2
 -\frac{2\lambda^2+\lambda\mu+\mu^2}{2(\lambda+\mu)^3}AB-\frac{\mu^2}{2(\lambda+\mu)^3}(A+2B)C-\frac{3\lambda\mu}{2(\lambda+\mu)^2}E\notag\\[1ex]
 &+&\frac{\mu^2}{(\lambda+\mu)^2}F+\left[2+\frac{\lambda^2}{(\lambda+\mu)^2}\right]G,
 \end{eqnarray}
 are the `large acoustoelasticity' coefficients.

We may also find expressions for the acoustoelastic effect in terms of the pre-stress $\sigma_1$ as
\begin{equation} \label{uni-s}
\rho v_{ii}^2 =  \lambda + 2 \mu + c_{ii}\sigma_1 + d_{ii} \sigma_1^2,\quad \rho v_{ik}^2 =  \mu + c_{ik} \sigma_1 + d_{ik} \sigma_1^2,
\end{equation}
where $i=1,2$ (no sum on repeated $i$), and $k=1,3$ ($k \ne i$).
Using \eqref{elong1}, we find that
\begin{equation}
c_{ij} =  \left(\dfrac{\lambda+\mu}{3 \kappa\mu}\right)a_{ij}, \quad
d_{ij} =  \left(\dfrac{\lambda+\mu}{3 \kappa\mu}\right)^2 \left(b_{ij}-\dfrac{\lambda+\mu}{3 \kappa\mu}\gamma a_{ij} \right),
\end{equation}
where $ij \in \{11, 22, 12, 21, 23\}$.
In particular, we recover
\begin{eqnarray}
c_{11} &=&  \dfrac{1}{3\kappa}\left[\lambda + 2B + 2C + 2\dfrac{\lambda + \mu}{\mu}\left(2\lambda+5\mu+A+2B\right) \right],
\notag \\[1ex]
c_{12} &=& \dfrac{1}{3\kappa}\left[4(\lambda+\mu) + \dfrac{\lambda+2\mu}{4\mu} A +B \right],
\notag \\[1ex]
c_{22} &=&  \dfrac{2}{3\kappa}\left[B + C - \dfrac{\lambda}{\mu}\left(\lambda+2\mu+\dfrac{A}{2}+B\right) \right],
\notag \\[1ex]
c_{21} &=& \dfrac{1}{3\kappa}\left[\lambda + 2\mu + \dfrac{\lambda+2\mu}{4\mu}A + B \right],
\notag\\[1ex]
c_{23} &=& - \dfrac{1}{3\kappa}\left[2\lambda + \dfrac{\lambda}{2\mu}A - B \right],
\end{eqnarray}
for the classical acoustoelastic effect \citep{HuKe53}.
Note that $c_{12} -c_{21}=1$ and that $d_{12} - d_{21} = 1/(3\kappa)$, and hence in the incompressible limit $d_{12}=d_{21}$.
Using these relations, we establish that 
\begin{equation}
\rho (v_{12}^2 - v_{21}^2)  = \sigma_1 + 1/(3\kappa) \sigma_1^2.\label{v2diff}
\end{equation}
Recall now that $\rho$ is the mass density in the reference configuration. When expressed in terms of the deformed density $\rho_c = \rho J^{-1}$ equation \eqref{v2diff} becomes
\begin{equation}
\rho_c (v_{12}^2 - v_{21}^2)  = \sigma_1.
\end{equation}
This relation is in fact exact in accordance with \eqref{calAdifference} and the expressions $\rho v_{ij}^2=J\mathcal{A}_{0ijij}$, $i\neq j$ specialized accordingly with $\sigma_2=0$.

Finally, we take the incompressible limits of the elastic constants using the limiting values listed in Section \ref{Introduction}.
There, we have $\rho v_{11}^2 \rightarrow \infty$ and $\rho v_{22}^2 \rightarrow \infty$, unsurprisingly, because longitudinal homogeneous plane waves may not propagate in incompressible solids.
For the transverse principal waves travelling in the direction of tension,
\begin{equation} \label{uni-e-incompr}
\rho v_{12}^2 =  \mu +\left(3\mu+\frac{A}{4} \right) e_1 + \left(5\mu+\frac{7}{4} A+ 3 D \right) e_1^2,
\end{equation}
in terms of the elongation  \citep{Destrade10}, and
\begin{equation} \label{uni-s-incompr}
\rho v_{12}^2 =  \mu + \left(1 +\dfrac{A}{12\mu}\right) \sigma_1 + \frac{1}{9\mu^2} \left[ 2 \mu + \frac{A}{4} \left( 3- \frac{A}{4\mu} \right) + 3D \right] \sigma_1^2,
\end{equation}
in terms of the pre-stress.
For the transverse principal waves travelling perpendicular to the direction of tension,
\begin{align} \label{uni-e-incompr2}
&\rho v_{21}^2 =  \mu + \frac{A}{4} e_1 + \left(2\mu+A+3D \right) e_1^2,\notag \\
&\rho v_{23}^2 =  \mu - \left(3\mu + \frac{A}{2} \right) e_1 + \left(5\mu+\frac{7}{4} A+3D \right) e_1^2,
\end{align}
in terms of the elongation, see \citep{Destrade10}, and, using  \eqref{sig},
\begin{align} \label{uni-s-incompr2}
&\rho v_{21}^2 =  \mu  + \dfrac{A}{12\mu} \sigma_1 +\frac{1}{9\mu^2} \left[ 2 \mu + \frac{A}{4} \left( 3- \frac{A}{4\mu} \right) + 3D \right] \sigma_1^2,\notag \\
&\rho v_{23}^2 =  \mu - \left(1 +\dfrac{A}{6 \mu}\right) \sigma_1 +  \frac{1}{9\mu^2} \left( 8 \mu + 3A + \frac{A^2}{8\mu} + 3D \right) \sigma_1^2,
\end{align}
in terms of the pre-stress.

\begin{figure}[!t] 
\centering
\epsfig{figure=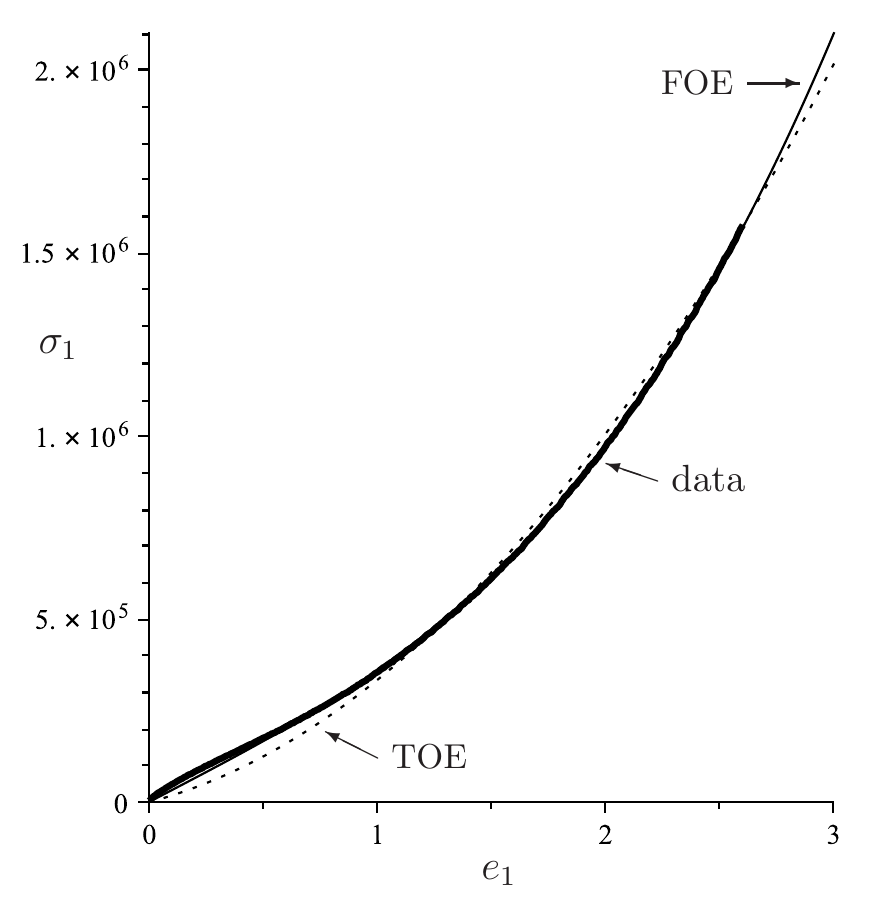,width=.5\textwidth}
\caption{Fitting tensile data (thick curve) of the Cauchy stress $\sigma_1$ (in Pa) versus the elongation $e_1$ for a sample of incompressible silicone rubber; the TOE strain-energy density cannot capture the variations adequately; the FOE model gives an excellent fit with only 3 material parameters, with values $\mu = 109.35$ kPa, $A = -454.18$ kPa, $D= 109.27$ kPa.}\label{fig:silicone1}
\end{figure}

For an application, we use data for a sample of silicone rubber that has been subjected to a standard tensile test. Figure \ref{fig:silicone1} displays the variation of the tensile Cauchy stress component $\sigma_1$ with the elongation up to a maximum stretch of about 250\%, at which stage one end of the sample snapped out of its grip. Over that range, the TOE strain energy density \eqref{energytoe} is not able to capture the behaviour of the sample adequately, as shown in Fig. \ref{fig:silicone1}, and is thus discarded. On the other hand, the FOE strain energy \eqref{energyfoe} gives an excellent least-squares fit, with  coefficient of correlation $R^2 = 0.9997$.  In fact, the fit is very good up to a much larger value of the stretch than could be expected of this fourth-order approximate theory.
We determined the following values for the constants of second-, third-, and fourth-order elasticity:
\begin{equation}
\mu = 109.35 \text{ kPa}, \quad A = -454.18 \text{ kPa}, \quad D = 109.27 \text{ kPa}.
\end{equation}
Note that all three are of the same order of magnitude, as expected from the theory \citep{DeOg10}.
We also remark that $-8\mu<A<-4\mu$, indicating that the corresponding Mooney--Rivlin solid is materially stable \citep{DeOg10,DeGM09}.

\begin{figure}[!h] 
\centering
\epsfig{figure=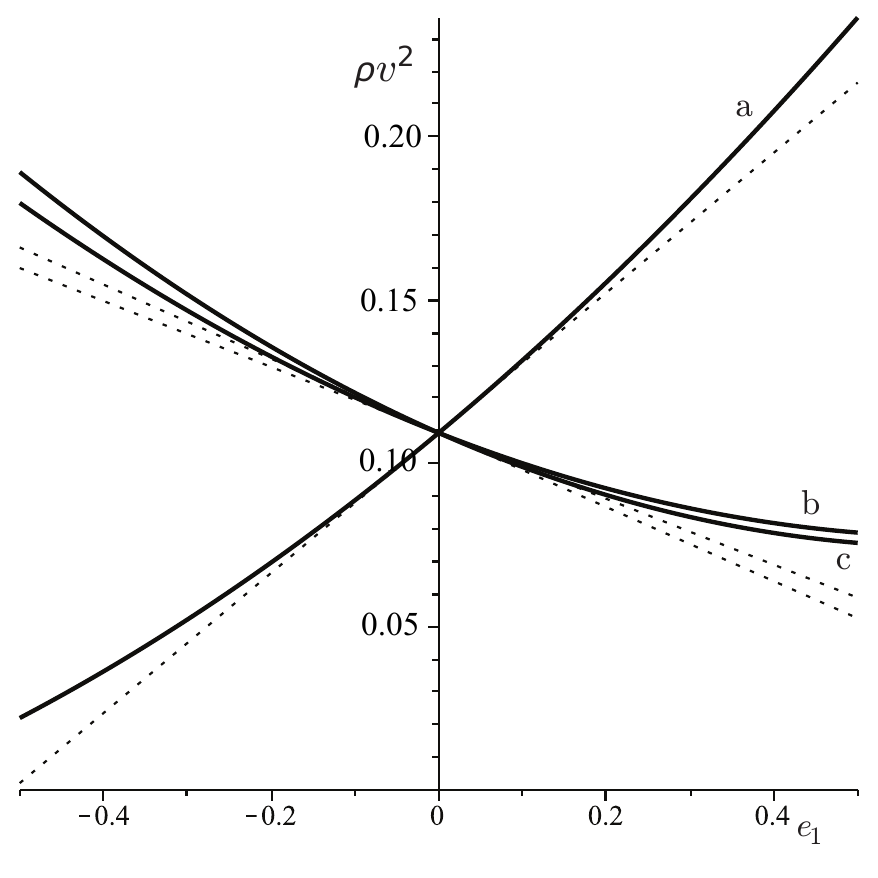, width=.5\textwidth}
\caption{Plots of $\rho v^2$ (in MPa) versus elongation $e_1$ for a deformed incompressible block of silicone rubber subject to simple tension; theoretical curves for waves travelling (a) in the direction $x_1$ of tension ($v=v_{12}$), (b) in any direction in the plane normal to the direction of uniaxial stress, with transverse polarization in that plane ($v=v_{23}$), (c) in any direction in the plane normal to the direction of uniaxial stress, with polarization normal to that plane ($v=v_{21}$); the dotted lines correspond to the linear part of the $\rho v^2$ versus $e_1$ curves (the classical acoustoelastic effect). }\label{fig:silicone2}
\end{figure}

Using these values, the theoretical variations of the squared wave speeds $ \rho v^2_{12}$, $\rho v^2_{21}$,  $\rho v^2_{23}$ are plotted versus the elongation in Fig. \ref{fig:silicone2}. These curves suggest that it is sufficient to elongate the sample by about 20\% to reveal quadratic acoustoelastic effects, and thus to determine $D$ experimentally.

\section{Discussion\label{discussion}}


This study was set within the framework of the theory of small-amplitude waves propagating in a deformed elastic solid. Existing general expressions for the speeds of body waves in a finitely deformed isotropic elastic solid were recalled and then specialized to fourth-order elasticity. Specifically, the squared wave speeds were expanded in terms of the pre-deformation (and, equivalently, in terms of the pre-stress) in order to reveal the so-called acoustoelastic effect at the considered order. While the classical acoustoelastic effect is concerned with the linear variation of the squared wave speed with the pre-stress, the expansion was extended to the quadratic regime (large acoustoelastic effect). Explicit expressions for the quadratic coefficients for compressible and incompressible materials were determined for the cases of hydrostatic and uniaxial pre-stress. These coefficients give access to the fourth-order elastic constants.

In the case of an incompressible sample under uniaxial tension, for example, measurement of the variations of the speed of a single transverse wave with respect to the pre-stress (or pre-strain) is enough to determine the elastic constants.  The wave speed in the unstressed configuration gives the initial shear modulus $\mu$ of second-order elasticity, the linear variation of the squared wave speed with the pre-stress or pre-strain gives the third-order Landau-coefficient $A$, and the quadratic variation gives the fourth-order constant $D$.

For compressible materials,   \cite{HuKe53} showed that any three of the seven equations \eqref{hydro-e}, \eqref{uni-e-1}, \eqref{uni-e-2}, with just the linear terms retained, or equivalently \eqref{hydro-s}, \eqref{uni-s}, excluding either the expression for $\rho v_{12}^2$ or for $\rho v_{21}^2$, suffice  to determine the two Lam\'e coefficients $\lambda$ and $\mu$ of second-order elasticity and the three Landau coefficients of third-order elasticity $A$, $B$, $C$. 
Here we have shown that any four of these seven equations, with the quadratic terms now included, give access also to the fourth-order constants $E$, $F$, $G$, $H$ (again, excluding either the expression for $\rho v_{12}^2$ or for $\rho v_{21}^2$).

Although we have restricted attention largely to third- and fourth-order elasticity the expressions for the components of the tensor of instantaneous elastic moduli $\bm{\mathcal{A}}_{0}$ given in \eqref{Aoijkl} apply for an arbitrary finite deformation relative to an unstressed configuration of an isotropic elastic material, and the subsequent incremental response depends on the finite deformation and its accompanying pre-stress.  If instead there is an initial stress in the reference configuration then the components of $\bm{\mathcal{A}}_{0}$ are considerably more complicated, as detailed in \cite{Shams2011}, but they do clarify how the elastic moduli depend in general on the initial stress.  In particular, the dependence of $\bm{\mathcal{A}}_{0}$ on an initial hydrostatic stress $\tau$ can be put in the simple form 
\begin{equation}
\mathcal{A}_{0ijkl}=\mu(\tau)(\delta_{ik}\delta_{jl}+\delta_{il}\delta_{jk})+\lambda(\tau)\delta_{ij}\delta_{kl},
\end{equation}
where $\mu(\tau)$ and $\lambda(\tau)$ are now stress-dependent Lam\'e moduli, with $\tau>0\, (<0)$ corresponding to tension (pressure), their precise functional dependence determined by the choice of strain-energy function.
We refer to a recent paper by \cite{Raj2009} and references therein for a discussion of the stress dependence of moduli within the context of  an implicit theory of elasticity.


\section*{Acknowledgements}


This work was supported by Erasmus funding from the European Commission and travel funding from the University of Salento and from the National University of Ireland Galway.  
It was also supported by an International Joint Project grant from the Royal Society of London. 
We thank Stephen Kiernan and Michael Gilchrist at University College Dublin for assistance with the tensile test of a silicone sample.



\begin{thebibliography}{99}

\bibitem[Birch(1938)]{Birc38}
{F. Birch}:
{The effect of pressure upon the elastic parameters of isotropic solids, according to Murnaghan's theory of finite strain}, 
J. Appl. Phys. 
\textbf{9}, 279--288 (1938).

\bibitem[Bland(1969)]{Blan69}
{D.R. Bland},
{Nonlinear Dynamic Elasticity},
Blaisdell, Waltham (1969).

\bibitem[Bridgman(1945)]{Brid45}
{P.W. Bridgman},
{The Physics of High Pressures}, Bell \& Sons, London (1945).

\bibitem[Brillouin(1925)]{Bril25}
{L. Brillouin}
{Sur les tensions de radiation}, Ann. Phys. ser. 10 \textbf{4}, 528--586  (1925).
	
\bibitem[Destrade et al.(2010a)]{Destrade10}
M. Destrade, M.D. Gilchrist, G. Saccomandi,
Third- and fourth-order constants of incompressible soft solids
and the acousto-elastic effect,
J. Acoust. Soc. Am. \textbf{127},  2759--2763 (2010).

\bibitem[Destrade et al.(2010b)]{DeGM09}
{M. Destrade, M.D. Gilchrist, and J.G. Murphy},
{Onset of non-linearity in the elastic bending of blocks},
ASME J. Appl. Mech.
\textbf{77}, 061015 (2010).

\bibitem[Destrade and Ogden(2010)]{DeOg10}
{M. Destrade and R.W. Ogden},
{On the third- and fourth-order constants of incompressible isotropic elasticity},
J. Acoust. Soc. Am.
\textbf{128}, 3334--3343 (2010).

\bibitem[Hamilton et al.(2004)]{HaIZ04}
{M. F. Hamilton, Y.A. Ilinskii, and E.A. Zabolotskaya},
{Separation of compressibility and shear deformation in the elastic
  energy density},
J. Acoust. Soc. Am.
\textbf{116}, 41--44 (2004).

\bibitem[Gennisson et al.(2007)]{Genn07}
{J.-L. Gennisson, M. R\'enier, S. Catheline, C. Barri\`ere, J. Bercoff, M. Tanter, and M. Fink},
{Acoustoelasticity in soft solids: Assessment of the nonlinear shear modulus with the acoustic radiation force},
J. Acoust. Soc. Am.
\textbf{122}, 3211--3219 (2007).

\bibitem[Hughes and Kelly(1953)]{HuKe53}
{D.S. Hughes and J.L. Kelly}
{Second-order elastic deformation of solids},
Phys. Rev.
\textbf{92}, 1145--1149 (1953).

\bibitem[Jacob et al.(2007)]{JCGB07}
{X. Jacob, S. Catheline, J.-L. Gennisson, C. Barri\`ere, D. Royer, and M. Fink},
{Nonlinear shear wave interaction in soft solids},
J. Acoust. Soc. Am.
\textbf{122}, 1917--1926 (2007).

\bibitem[Kim and Sachse(2001)]{KiSa01}
{K.Y. Kim and W. Sachse}
{Acoustoelasticity of elastic solids},
in {Handbook of Elastic Properties of Solids, Liquids, and Gases},
Levy, Bass, Stern (Editors),
\textbf{1}, 441--468, Academic Press, New York, (2001).

\bibitem[Landau and Lifshitz(1986)]{LaLi86}
{L.D. Landau and E.M. Lifshitz}
{Theory of Elasticity}, 3rd ed. Pergamon, New York (1986).

\bibitem[Mironov et al.(2009)]{Miro09}
{M.A. Mironov, P.A. Pyatakov, I.I. Konopatskaya, G.T. Clement, and N.I. Vykhodtseva},
{Parametric excitation of shear waves in soft solids},
Acoust. Phys. \textbf{55}, 567--574 (2009).

\bibitem[Murnaghan(1951)]{Murn51}
{F.D. Murnaghan},
{Finite Deformation of an Elastic Solid}, Wiley, New York (1951).

\bibitem[Ogden(1974)]{Ogde74}
{R.W. Ogden},
{On isotropic tensors and elastic moduli},
Proc. Cambr. Phil. Soc.
\textbf{75}, 427--436  (1974).

\bibitem[Ogden(1997)]{Ogde97}
{R.W. Ogden},
{Non-linear Elastic Deformations}. Dover, New York (1997).

\bibitem[Ogden(2007)]{Ogde07}
{R.W. Ogden},
{Incremental statics and dynamics of pre-stressed elastic materials},
in {Waves in Nonlinear Pre-Stressed Materials},
M. Destrade, G. Saccomandi (Editors),
CISM Lecture Notes,
\textbf{495}, 1--26. Springer, New York  (2007).

\bibitem[Pao et al.(1984)]{PaSa84}
{Y.-H. Pao, W. Sachse,  H. Fukuoka},
{Acoustoelasticity and ultrasonic measurements of residual stresses}. 
In W.P. Mason and R.N. Thurston, editors, \emph{Physical Acoustics}, 
\textbf{17}, 61--143. Academic Press, New York (1984).

\bibitem[R\'enier et al.(2007)]{Reni07}
{M. R\'enier, J.-L. Gennisson, M. Tanter,  S. Catheline, C. Barri\`ere, D. Royer, and M. Fink},
{Nonlinear shear elastic moduli in quasi-incompressible soft solids},
IEEE Ultras. Symp. Proc, 554--557 (2007).

\bibitem[R\'enier et al.(2008a)]{Reni08a}
{M. Renier, J.-L. Gennisson, C. Barriere, S. Catheline, M. Tanter, D. Royer, and M. Fink},
{Measurement of shear elastic moduli in quasi-incompressible soft solids},
18th International Symposium on Nonlinear Acoustics, July 07-10, 2008 Stockholm, Sweden,
{Nonlinear Acoustics Fundamentals and Applications},
Book Series: AIP Conference Proceedings,
\textbf{1022}, 303--306  (2008).

\bibitem[R\'enier et al.(2008b)]{Reni08b}
{M. R\'enier, J.-L. Gennisson, C. Barri\`ere, D. Royer, and M. Fink},
{Fourth-order shear elastic constant assessment in quasi-incompressible soft solids},
Appl. Phys. Lett.
\textbf{93}, 101912  (2008).

\bibitem[Tang(1967)]{Tang67}
{S. Tang},
{Wave propagation in initially-stressed elastic solids}, 
Acta Mech. 
\textbf{4}, 92--106 (1967).

\bibitem[Rajagopal and Saccomandi(2009)]{Raj2009}
{K.R. Rajagopal and G. Saccomandi},
{The mechanics and mathematics of the effect of pressure on the shear modulus of elastomers},
Proc. R. Soc. Lond. A \textbf{465}, 3859--3874 (2009).

\bibitem[Shams et al.(2011)]{Shams2011}
{M. Shams, M. Destrade, and R.W. Ogden}, {Initial stresses in elastic solids: Constitutive
laws and acoustoelasticity}, Wave Motion \textbf{48}, in press. DOI:10.1016/j.wavemoti.2011.04.004

\bibitem[Zabolotskaya et al.(2004)]{HaIZ04B}
{ E.A. Zabolotskaya,  Y.A. Ilinskii, M. F. Hamilton, and G. D. Meegan},
{Modeling of nonlinear shear waves in soft solids},
J. Acoust. Soc. Am. \textbf{116}, 2807--2813 (2004).

\bibitem[Zabolotskaya et al.(2007)]{HaIZ07}
{ E.A. Zabolotskaya,  Y.A. Ilinskii, and M. F. Hamilton},
{Nonlinear surface waves in soft, weakly compressible elastic media},
J. Acoust. Soc. Am.
\textbf{121}, 1873--1878 (2007).

	
\end{thebibliography}
\end{document}